\documentclass[12pt]{iopart}
\usepackage{iopams}
\usepackage{graphicx}

\newcommand{\eqref}[1]{(\ref{#1})}
\newcommand{\commento}[1]{}

\newcommand{\be}{\begin{equation}}
\newcommand{\ee}{\end{equation}}
\newcommand{\bea}{\begin{eqnarray}}
\newcommand{\eea}{\end{eqnarray}}
\newcommand{\barr}{\begin{array}}
\newcommand{\earr}{\end{array}}

\newcommand{\ep}{\epsilon}
\newcommand{\sen}{\sin\theta}
\newcommand{\cs}{\cos\theta}
\newcommand{\al}{\alpha}

\newcommand{\sens}{\sin^2\theta}
\newcommand{\css}{\cos^2\theta}
\newcommand{\senf}{\sin^4\theta}
\newcommand{\sensix}{\sin^6\theta}

\graphicspath{{Figures/}}

\begin{document}

\title[Models of granular ratchets]{Models of granular ratchets}

\author{G. Costantini$^{1,2}$, A. Puglisi$^{1,3}$ and U. Marini Bettolo Marconi$^2$}
\address{$^1$ CNR - INFM Statistical Mechanics and Complexity Center, p.le Aldo Moro 7, 00185 Roma, Italy}
\address{$^2$ Universit\`a di Camerino, Dipartimento di Fisica, Via Madonna delle Carceri, 62032 Camerino, Italy}
\address{$^3$ Dipartimento di Fisica, Universit\`a La Sapienza, p.le Aldo Moro 2, 00185 Roma, Italy}

\ead{giulio.costantini@unicam.it}

\begin{abstract}
  We study a general model of granular Brownian ratchet consisting of
  an asymmetric object moving on a line and surrounded by a
  two-dimensional granular gas, which in turn is coupled to an
  external random driving force. We discuss the two resulting
  Boltzmann equations describing the gas and the object in the dilute
  limit and obtain a closed system
  for the first few moments of the system velocity
  distributions. Predictions for the net ratchet drift, the variance
  of its velocity fluctuations and the transition rates in the
  Markovian limit, are compared to numerical simulations and a fair
  agreement is observed.
\end{abstract}

\maketitle

\section{Introduction}

A Brownian ratchet is a system designed to extract work, usually in
the form of a net drift or current, from a thermal bath. If the bath
is at equilibrium, i.e. it is characterized by only one temperature, a
ratcheting behavior is prevented by the second principle of
thermodynamics. The system must be coupled to different baths at
different temperatures, or to additional specific non-conservative
(e.g. time-dependent) forces, in order to escape the consequences of
the second principle. Moreover, to observe a net drift, spatial
symmetry must also be broken~\cite{hanggi,vdb2,vdb3}.  Dissipation of energy in
inelastic collisions between macroscopic grains~\cite{gases}, breaks the
time-reversal symmetry and leads to the introduction of suggestively
simple models of inelastic Brownian ratchets, which are apparently
coupled to only one thermal bath at a single temperature. It is known,
for instance, that a granular object, surrounded by a stationary
inelastic gas and characterized by a left-right asymmetry, presents a
rectification of thermal fluctuations \cite{noi} \cite{vdb1}
resulting in a net drift in a given direction: the
asymmetry of the object can originate from its shape or its
inelasticity profile (i.e. different inelasticities on different
portions of the surface). Interestingly, the probability density
function (pdf) $P(V)$ of the object velocity results asymmetric when
its mass $M$ is of the same order or smaller than that of surrounding
disks~\cite{epl}: such an effect is stronger the smaller the
elasticity.  In this study we intend to offer an analysis of a general
model which includes all ingredients cited above, treating also the
dynamics of the surrounding gas. This will make clear the conditions
required to decouple the gas dynamics from that of the object, making
the latter obey a closed Markovian master equation. In this limit we
will obtain some general formula for the net drift of the object, its
velocity variance and the transition rates of the Markov process,
which in general do not satisfy detailed balance. These results
compare very well with numerical simulations.  The present article is
organized as follows: in Sec.\ref{Theory} we introduce the model and
obtain the coupled equations of evolution for the probability
distributions of the velocities of the ratchet and of the gas, in
Sec.\ref{Evolution} instead of solving directly the former equations
we consider the governing equations for the moments of the
distributions, while in Sec.~\ref{Simple} we specialize our study to
the case of an equilateral triangle, while in Sec~\ref{Conclusion} we
conclude with a brief summary.  Finally, we provide an appendix
containing the necessary formulas for the coefficients entering the
equations of Sec.\ref{Evolution}

\section{Theory}
\label{Theory}

Our 2D model consists of a rigid convex object of mass $M$, generally
asymmetric, surrounded by a dilute gas of $N$ hard disks of mass $m$
and density $n=N/A$ where $A$ is the area of the box. The surface of the object, of
perimeter $C$, has a non-homogeneous inelasticity with a coefficient
of restitution that depends on the point of contact on the surface.
The object can only slide, without rotating, along the direction
$x$. The collisions between two disks of the gas are dissipative with
a coefficient of restitution $\alpha_{dd}$.  In this system the energy
is not conserved and an external driving mechanism is needed to attain
a stationary state.  In order to maintain this steady state the gas is
coupled to a thermal bath~\cite{kicks1}: the gas particles between two collisions
are subject to an external random force. The dynamics of the object is
assumed not to couple directly with the thermostat, but only with the
gas particles. For the sake of simplicity we assume that the gas is
dilute and that Molecular Chaos is valid for object-disks collisions:
this allow us to use the Direct Simulation Monte-Carlo (DSMC)
algorithm to simulate the system dynamics~\cite{Bird}.


After a binary collision, the velocities of the particles and the
object, $\vec{v}$ and $\vec{V}$ respectively, can be obtained, from
their pre-collisional values $\vec{v}'=(v'_x,v'_y)$ and
$\vec{V}'=(V',0)$, imposing the following conditions
concerning a portion of the surface:
\begin{eqnarray}
\label{eq:conditions1}
MV+mv_x&=& MV'+mv'_x \\
\label{eq:conditions2}
\vec{v} \cdot\hat{k}_{\parallel}&=& \vec{v}'\cdot\hat{k}_{\parallel}  \\
\label{eq:conditions3}
(\vec{V}-\vec{v})\cdot \hat{k}_{\perp} &=& -\alpha(\theta)(\vec{V}'-\vec{v}')\cdot \hat{k}_{\perp}.
\end{eqnarray}
$\hat{k}_{\parallel}$ and $\hat{k}_{\perp}$ are the unit vectors,
parallel and perpendicular respectively, to the object surface in the
collision point (see Fig. \ref{triangle}). These can be
expressed as $\hat{k}_{\parallel}=(\cos \theta,\sin\theta)$ and
$\hat{k}_{\parallel}=(\sin \theta,-\cos\theta)$ where $\theta$ is the
angle created by $\hat{k}_{\parallel}$ and the $x$ axes, modulus
$2\pi$ (Fig. \ref{triangle}). In this way the coefficient of restitution is a function of $\theta$ and it is expressed as $\al(\theta)$. 
Equations (\ref{eq:conditions1}) and
(\ref{eq:conditions2}) correspond to momentum conservation in the $x$
direction and parallel to the surface, while the inelasticity takes
part only in the reduction of the relative velocity expressed in
Eq. (\ref{eq:conditions3}). Solving these equations, we can write the
post-collisional velocities as
\begin{eqnarray}
\label{eq:rules}
V &=& V'+\frac{[1+\al(\theta)]\ep^2\sin\theta}{1+\ep^2\sin^2\theta}[(v'_x-V')\sin\theta-v'_y\cos\theta] \nonumber \\
v_x &=&  v'_x-\frac{[1+\al(\theta)]\sin\theta}{1+\ep^2\sin^2\theta}[(v'_x-V')\sin\theta-v'_y \cos\theta] \nonumber \\
v_y &=& v'_y+\frac{[1+\al(\theta)]\cos\theta}{1+\ep^2\sin^2\theta}[(v'_x-V')\sin\theta-v'_y \cos\theta]
\end{eqnarray}
where $\ep^2=m/M$ is the mass ratio. The collision between two disks labeled $1$ and $2$, instead, is given by
\begin{eqnarray}
\label{eq:rulesdd}
\vec{v}_1 &=& \vec{v}'_1-\frac{1+\alpha_{dd}}{2}(\vec{v}_{12}\cdot \hat{n})\hat{n} \nonumber \\ 
\vec{v}_2 &=& \vec{v}'_2+\frac{1+\alpha_{dd}}{2}(\vec{v}_{12}\cdot \hat{n})\hat{n},
\end{eqnarray}
with $\vec{v}_{12}=\vec{v}_1-\vec{v}_2$ and $\hat{n}$ the unit vector in the direction joining the centers of the two disks.
As external driving force acting on the disks, we choose the following heat bath:
\be
m \frac{\partial \vec{v}}{\partial t}= -m \Gamma \vec{v}(t)+\sqrt{2m\Gamma T_b}~ \vec{\zeta}(t)
\label{bath}
\ee 
with $\vec{\zeta}(t)$ a Gaussian white noise with $\langle
\vec{\zeta} (t) \rangle=0$ and $\langle \zeta_i(t)
\zeta_j(t')\rangle=\delta_{ij}\delta(t-t')$ and $\Gamma\equiv
1/\tau_b$ the drag coefficient.  The model has been studied in
$2D$~\cite{kicks1} and $1D$~\cite{kicks2} and also without
viscosity~\cite{kicks3}. In the dilute gas limit, it is possible to
describe the object dynamics by means of a Boltzmann Equation (BE) for
$P(V,t)$ which can be written, if the object is convex, as \cite{noi}
\be
\label{boleq} 
\frac{\partial P(V,t)}{\partial
t}=\int dV'~[W(V|V')P(V',t)-W(V'|V)P(V,t)] 
\ee 
where the rates for the collision object-disk is 
\begin{eqnarray} 
W(V|V')= & \int_0^{2\pi}\!\!\!\!\!n\tilde{C} F(\theta) d\theta\int_{-\infty}^{+\infty} \!\!\!\!\!\!dv'_x 
\int_{-\infty}^{+\infty} \!\!\!\!\!\!dv'_y   \phi(v'_x,v'_y,t) \nonumber \\ 
&\Theta[(\vec{V}'-\vec{v}')\cdot \hat{k}_{\perp}] 
(\vec{V}'-\vec{v})\cdot \hat{k}_{\perp}  
\delta \Big \{ V-V'\nonumber \\
&- \frac{\sen}{\kappa(\theta,\al(\theta))} \left[(v'_x-V')\sin\theta-v'_y\cos\theta \right] \Big\}
\label{trate}
\end{eqnarray}
with $\kappa(\theta,\al(\theta))=(1+\ep^2\sin^2\theta)/[(1+\al(\theta))\ep^2]$, $\Theta$ is the Heaviside step function
and $\phi(v_x,v_y,t)$ is
the pdf of the gas particles.  The $F(\theta)$ in Eq. \eqref{trate} is
a shape function of the object and it is such that
$dl=\tilde{C}F(\theta)d\theta$ is the length of its outer
''effective'' surface $\tilde{C}$ that has a tangent between $\theta$
and $\theta+d\theta$ (see Fig.~\ref{triangle} for an example). 

Also the dynamics of the pdf, $\phi(\vec{v},t)$, of the gas particles
obeys, in the dilute limit, a BE analogous to 
Eq. (\ref{boleq}), with the contributions of disk-disk
collisions,  disk-object collisions and the coupling with the external
driving. Then the BE for $\phi(\vec{v},t)$ can be written as
\begin{eqnarray}
\label{boleqgas} 
\frac{\partial \phi(\vec{v},t)}{\partial
t}=&J[\vec{v}|\phi,\phi]+\int dv'_x \int dv'_y ~[W(\vec{v}|\vec{v}')\phi(\vec{v}',t)-\nonumber \\ &W(\vec{v}'|\vec{v})\phi(\vec{v},t)]
+\mathcal B\phi(\vec{v},t).
\end{eqnarray}
In the above equation $J[\vec{v}|\phi,\phi]$ is the Boltzmann
collision operator for the disk-disk interactions \cite{garzo}, while
$\mathcal B$ is an operator representing the effects of the viscous
force and of an external bath allowing the granular gas to reach the
steady state. If we consider, as thermostat, the bath of Eq.~\eqref{bath},
this operator has the form 
\be \mathcal B \phi(\vec{v},t)=\Gamma
\frac{\partial}{\partial \vec{v}}\Big[\vec{v} \phi(\vec{v},t)\Big]+Q
\bigtriangleup_v [\phi(\vec{v},t)] 
\ee
where $\bigtriangleup_v$ is the Laplacian operator with respect to the velocity~\cite{risken}. \\
The transition rate in Eq. (\ref{boleqgas}) is given, instead, by
\begin{eqnarray} 
W(\vec{v}|\vec{v}')= & \int_0^{2\pi}\!\!\!\!\!n_{ob}\tilde{C}F(\theta) d\theta
\!\int_{-\infty}^{+\infty}\!\!\!\!\!\!\! dV 
P(V',t)(\vec{V}'-\vec{v}')\cdot \hat{k}_{\perp} \nonumber \\ 
&\Theta[(\vec{V}'-\vec{v}')\cdot \hat{k}_{\perp}] 
\delta \Big \{v_x-v'_x-\frac{\sin\theta}{\ep^2\kappa(\theta,\al(\theta))} \cdot \nonumber \\
&\big[\sin\theta(V'-v'_x) +v'_y\cos\theta \big]\Big\} 
\delta \Big \{ v_y-v'_y+\nonumber \\
& \frac{\cos\theta}{\ep^2\kappa(\theta,\al(\theta))}\cdot
\big[\sin\theta(V'-v'_x)  + v'_y\cos\theta \big]\Big\}
\label{trated}
\end{eqnarray}
where $n_{ob}=1/A$ is the density of objects in the system. The
function $\tilde{C}F(\theta)$ is the same of Eq. (\ref{trate}) because
it is connected to the differential cross section that is a property
of the colliding couple.\\ A first quantitative information about this
system is the collision frequency $\omega_c^{rd}$ between ratchet and
disks. It can be obtained from the transition rate (\ref{trate}),
using the relation \be \omega_c^{rd}=\int_{-\infty}^{+\infty} dV P(V)
\int_{-\infty}^{+\infty} W(V'|V) dV'.
\label{coll}
\ee 
The value of $\omega_c^{rd}$ is then determined by the choice of
$P(V)$.  In our previous article \cite{epl} we have shown that, if the
ratchet mass is comparable or smaller than the disk mass, i.e.
$\ep^2\gtrsim 1$, the ratchet pdf is asymmetric and deviates strongly
from a Maxwellian distribution. It is essential, in this case, to
include also the third moment of the distribution~\cite{Sela}. Then we assume,
for the object, a $P(V)$ of the form 
\be 
P(V)=\sqrt{\frac{M}{2\pi T_r}}\left(1-\frac{\xi}{6} \frac{\partial^3}{\partial V^3}\right)
\exp \Big[-\frac{M(V-\langle V\rangle)^2}{2T_r}\Big]
\label{pdfnogauss}
\ee
where $T_r=M\langle (V-\langle V \rangle)^2 \rangle$ is the granular temperature of the ratchet and $\xi=\langle (V-\langle V \rangle)^3\rangle$ is a
measure of the asymmetry of $P(V)$ about the average value.
Assuming that $\langle
V \rangle \ll V_{th}\equiv\sqrt{2T_r/M}$ and retaining only
the terms of first order in $\langle V \rangle $, we can perform the integrations in Eq. (\ref{coll}) obtaining that
\begin{eqnarray} 
\omega_c^{rd}=&\frac{1}{\sqrt{\pi}\tau^0_{rd}} \Big[
 \frac{u_1(z)}{2}-
\frac{u_2(z)}{3V^4_{th}}\xi\left( \langle V \rangle -\langle v_x \rangle \right)  - \frac{u_3(z)}{3V^4_{th}}\xi \langle v_y \rangle \Big]
\label{collf}
\end{eqnarray} 
where $\tau^0_{rd}=(n\tilde{C} V_{th})^{-1}$, $z=1/(\eta\ep^2)$ and
$\eta=T_r/T_g$ is the ratio between the temperature of the tracer and
the gas. Obviously $\xi$ must be small enough to make positive formula~\ref{collf}, otherwise the assumptions must be revisited.  The coefficients in Eq. (\ref{collf}), which depend on $z$
and on the shape of the object, are given explicitly in the Appendix.\\

\section{Evolution of moments}
\label{Evolution}
The mean velocity and granular temperature of the object, as well as
those of the gas, can be calculated from the two BE above,
Eqs.~\eqref{boleq} and~\eqref{boleqgas}. Starting from these, in fact, we
obtain a set of equations for the first moments of the distributions,
using suitable approximations to close the set.  To this aim we can assume,
reasonably, a Maxwellian distribution of the disk velocities since the
gas is directly coupled to the bath, i.e. $\phi(\vec{v})=m/(2\pi
T_g)\exp[-m(\vec{v}-\langle\vec{v}\rangle)^2/(2T_g)]$. The granular
temperature of the gas is therefore $T_g=m \langle
(\vec{v}-\langle\vec{v}\rangle)^2 \rangle/2$.  The equations for the
first three moments of the distribution of $P(V)$ can be obtained
multiplying both sides of~\eqref{boleq} by $V$, $M(V-\langle
V\rangle)^2$ and $(V-\langle V\rangle)^3$ respectively, and performing
the integrations.\\
The analogous equations for $\langle \vec{v} \rangle$ and $T_g$ can be
extracted in the same way, starting from Eq. (\ref{boleqgas}) and
considering also the contribution deriving from disk-disk collisions.\\
After long calculations, with the further assumption $\langle V
\rangle \ll V_{th}\equiv\sqrt{2T_r/M}$, we have derived the following
equations for the moments of the object
\begin{eqnarray}
\label{avequations1}
\frac{\partial \langle V \rangle}{\partial t} =&{}
-\frac{\ep^2}{\tau^0_{rd}}\Big[
\frac{V_{th}}{4}a_1(z) + \frac{a_2(z)}{\sqrt{\pi}}
 \left( \langle V\rangle - \langle v_x\rangle \right) + \nonumber \\
&\frac{a_3(z)}{\sqrt{\pi}}\langle v_y\rangle +
\frac{a_4(z)}{3\sqrt{\pi}V^2_{th}} \xi \Big ]\\
\label{avequations2}
\frac{\partial T_r}{\partial t} =&{}
\frac{M }{2\tau^0_{rd}}\Big[
\frac{V^2_{th}}{\sqrt{\pi}}  b_1(z) + 
\frac{V_{th}}{2} b_2(z)\left( \langle V\rangle -\langle v_x\rangle \right)+ \nonumber \\ 
& \frac{V_{th}}{2} b_3(z)\langle v_y\rangle + b_4 \xi+ 
\frac{b_5(z)}{3\sqrt{\pi}V^2_{th}} \xi \left( \langle V\rangle - \langle v_x\rangle \right)+\nonumber \\
&\frac{b_6(z)}{3\sqrt{\pi}V^2_{th}}  \langle v_y\rangle \xi \Big ]\\
\label{avequations3}
\frac{\partial \xi}{\partial t} =&{}
-\frac{1}{2\tau^0_{rd}}\Big[
\frac{3}{4}V^3_{th}c_1(z)+ \frac{V^2_{th}}{2\sqrt{\pi}} c_2(z) \langle V \rangle -\nonumber \\
&\frac{c_3(z)}{2\sqrt{\pi}} \xi -
\frac{c_4(z)}{V_{th}}  \langle V\rangle \xi - 
\frac{V^2_{th}}{2\sqrt{\pi}} c_5(z) \langle v_x\rangle +\nonumber \\
&\frac{V^2_{th}}{2\sqrt{\pi}} c_6(z) \langle v_y\rangle + 
\frac{c_7}{V_{th}} \langle v_x\rangle \xi - 
\frac{c_8}{V_{th}}  \langle v_y\rangle \xi  \Big ] 
\end{eqnarray}
and for those of the gas
\begin{eqnarray}
\label{avequations4}
\frac{\partial \langle v_x \rangle}{\partial t} =&{}
\frac{1}{N\tau^0_{rd}} \Big[
\frac{ V_{th}}{4} a_1(z)+\frac{a_2(z)}{\sqrt{\pi}}
\left( \langle V \rangle -\langle v_x \rangle \right) + \nonumber \\
&\frac{a_3(z)}{\sqrt{\pi}}\langle v_y \rangle + 
\frac{a_4(z)}{3\sqrt{\pi}V^2_{th}} \xi\Big ] -\Gamma \langle v_x \rangle \\
\label{avequations5}
\frac{\partial \langle v_y \rangle}{\partial t} =&{}
-\frac{1}{N\tau^0_{rd}} \Big[
\frac{V_{th}}{4}  d_1(z)+\frac{a_3(z)}{\sqrt{\pi}} \left( \langle V \rangle - \langle v_x \rangle \right) + \nonumber \\
&\frac{d_2(z)}{\sqrt{\pi}} \langle v_y \rangle + 
\frac{d_3(z)}{3\sqrt{\pi}V^2_{th}} \xi \Big ] -\Gamma \langle v_y \rangle \\
\label{avequations6}
\frac{\partial T_g}{\partial t} =&{}
\frac{m}{2N\tau^0_{rd}} \Big[
\frac{V^2_{th}}{2\sqrt{\pi}}  e_1(z) +
\frac{V_{th}e_2(z)}{4}  \left(\langle V \rangle- \langle v_x \rangle \right)+ \nonumber \\
&\frac{V_{th}}{4} e_3(z) \langle v_y \rangle + \frac{e_4}{2}  \xi +
\frac{e_5(z)}{3\sqrt{\pi}V^2_{th}} \left( \langle V \rangle- \langle v_x \rangle \right) \xi +\nonumber \\
&\frac{e_6(z)}{3\sqrt{\pi}V^2_{th}} \langle v_y \rangle \xi \Big ] - 
\frac{v^2_{th}}{2\tau_{dd}} (1-\alpha^2_{dd})- 2 \Gamma  v^2_{th} +4Q
\end{eqnarray}
where $v_{th}=\sqrt{2T_g/m}$ is the thermal velocity of the disks,
$\tau_{dd}=(\sqrt{2\pi}n\sigma v_{th})^{-1}$ is the collision time
between two disks and it is important to remind that $z$ is time
dependent. The coefficients of the above equations are given in the
Appendix.
Comparing the Eqs. (\ref{avequations1}) and (\ref{avequations4}) we
obtain that 
\be 
\frac{\partial \langle v_x \rangle}{\partial
  t}=-\frac{1}{N\ep^2 } \frac{\partial \langle V \rangle}{\partial t}-
\Gamma \langle v_x \rangle.  
\ee 
In the steady state this
implies that $\langle v_x \rangle= 0$. Moreover if the object is symmetric
(both in shape and inelasticity) with respect to the axis $x$, then the
coefficients $u_3$, $d_1$, $d_3$, $a_3$, $b_3$, $b_6$, $c_6$ and $c_8$
vanish (see Appendix). From Eq. (\ref{avequations5}) it results that
in this case the stationarity implies also $\langle v_y
\rangle=0$. 
The average velocity of the object, $\langle V \rangle$, in the steady state  is given by 
\be
\langle V
\rangle=-V_{th}\Big[\frac{\sqrt{\pi}}{4}\frac{a_1(z)}{a_2(z)}+3
\frac{\xi}{V^3_{th}}\frac{a_4(z)}{a_2(z)} \Big].
\ee 
As previously said the contribution of $\xi$ is decisive if $\ep^2\gtrsim1$, while it vanishes if the ratchet mass is very large respect to the disk mass. In
this case the average velocity is determined by the ratio
$a_1(z)/a_2(z)$. To give an example, if we consider an exact isosceles triangular
ratchet with angle opposite to the base equal to $2 \theta_0$ and with
$\al_1=\al_2=\al$, we have for $\ep^2\ll 1$
\begin{eqnarray} \label{isosc}
\langle V \rangle&=-\frac{\ep V_{th}\sqrt{\pi}}{4\sqrt{\eta}}\left(\eta-1\right) \frac{\int_0^{2\pi}d\theta F(\theta) \sin^3\theta}{\int_0^{2\pi}d\theta F(\theta) \sin^2\theta}=\nonumber \\
&=-\frac{1-\eta}{4}\sqrt{\frac{2\pi T_g}{m}}
\epsilon^2(1-\sin \theta_0).
\end{eqnarray}
This is just the equation for $\langle V \rangle$ obtained
in~\cite{noi}.  On the contrary, if a flat ``piston'' (perpendicular
to the $x$ axis) with the two faces with different inelasticities
$\alpha_1$, for the left face, and $\alpha_2$, for the right face, one
retrieves~\cite{epl}: \be \langle V \rangle=-\sqrt{\frac{2\pi
T_g}{m}}\frac{\al_2-\al_1}{4(2+\al_{dx}+\al_{sx})}\Big[ 1 +
\ep^2(\eta-1) \Big] \ee which, in general, has a larger signal-noise
ratio $\langle V \rangle/\sqrt{T_g}$, with respect to the uniformly inelastic case~\eqref{isosc}, and should be easier to be
observed in experiments.  In conclusion, the asymmetry of the system,
appearing in the coefficients $a_1$, $a_2$, and $a_4$, determines a
net drift of the object: a study of these coefficients shows that a
modulation of the inelasticity along the surface, i.e. a non-constant $\alpha(\theta)$
is more efficient in producing a net drift, with respect to the
geometrical asymmetry. It is also interesting, looking at
Eqs.~(\ref{avequations1})-(\ref{avequations6}), to discuss the degree
of coupling between the gas, the object and the thermal bath, which is
determined by the three characteristic times present in the system:
the relaxation time of the thermal bath $\tau_b$, the disk-disk
collision time $\tau_{dd}$ and the disk-ratchet collision time
$\tau^0_{dr}=N\tau^0_{rd}$. Here, we are interested in the case
$\tau_{dd} < \tau_b$, where inelastic
collisions among the gas particles become relevant; in this case, three scenarios can
occur: (i) when $\tau_{dr}<\tau_{dd}<\tau_b$, (ii) when
$\tau_{dd}<\tau_{dr}<\tau_b$ or (iii) when
$\tau_{dd}<\tau_b<\tau_{dr}$.  In case (i), the dynamics of the ratchet
and the disks are strongly coupled: in this case we expect the region
of gas surrounding the object to be correlated with the object itself,
making doubtful the assumptions of homogeneity and diluteness introduced
at the beginning to treat the system with Molecular Chaos.  If,
instead, the disk-disk collisions are more frequent, i.e. in cases
(ii) and (iii), fast and homogeneous relaxation of the gas is
expected. In particular, in  situation (iii) the gas dynamics is
dominated by internal collisions and by the external driving, and can
be regarded as uncoupled from the ratchet: in this limit the gas
velocity pdf $\phi(\vec{v})$ is constant and Eq.~\eqref{boleq} is a
Master Equation, i.e. the ratchet velocity performs a Markov
process~\cite{ap}. The condition for the occurrence of case (i) can be
approximated by \be
\frac{\tau_{dd}}{\tau^0_{dr}}=\frac{1}{N}\sqrt{\frac{\pi\eta
}{2}}\sqrt{\frac{\rho_d}{\rho_r}} \gg 1 \ee where
$\rho_d=4m/(\pi\sigma^2)$ and $\rho_r=4\pi M/\tilde{C}^2$ are
proportional to the mass densities of the disks and the ratchet,
respectively. Then case (i) occurs when $\rho_d/\rho_r>N^2$ and
this never takes place, in practice.  Cases (ii) and (iii), instead,
are characterized by the ratio $\tau^0_{dr}/\tau_b$ being smaller or
larger than $1$, respectively. Because $\tau^0_{dr} \propto
A/\tilde{C}$ and $A \gg \tilde{C}$, in order to neglect the excluded
area of the object, the case (ii) is obtained only if $\tau_b$ is
large enough. This implies very long simulations in order to obtain
average values statistically relevant. For this reason, we have
preferred to verify the Eqs.(\ref{avequations1})-(\ref{avequations6})
in situation (iii).  In principle, and in particular in cases (i) and
(ii), one should verify the stability of the stationary state. A study of linear stability is the objective of ongoing research and of a future
publication.

\section{A simple case: an equilateral triangle}
\label{Simple}

\begin{figure}[htbp]
\begin{center}
\includegraphics[angle=0,width=6cm,clip=true]{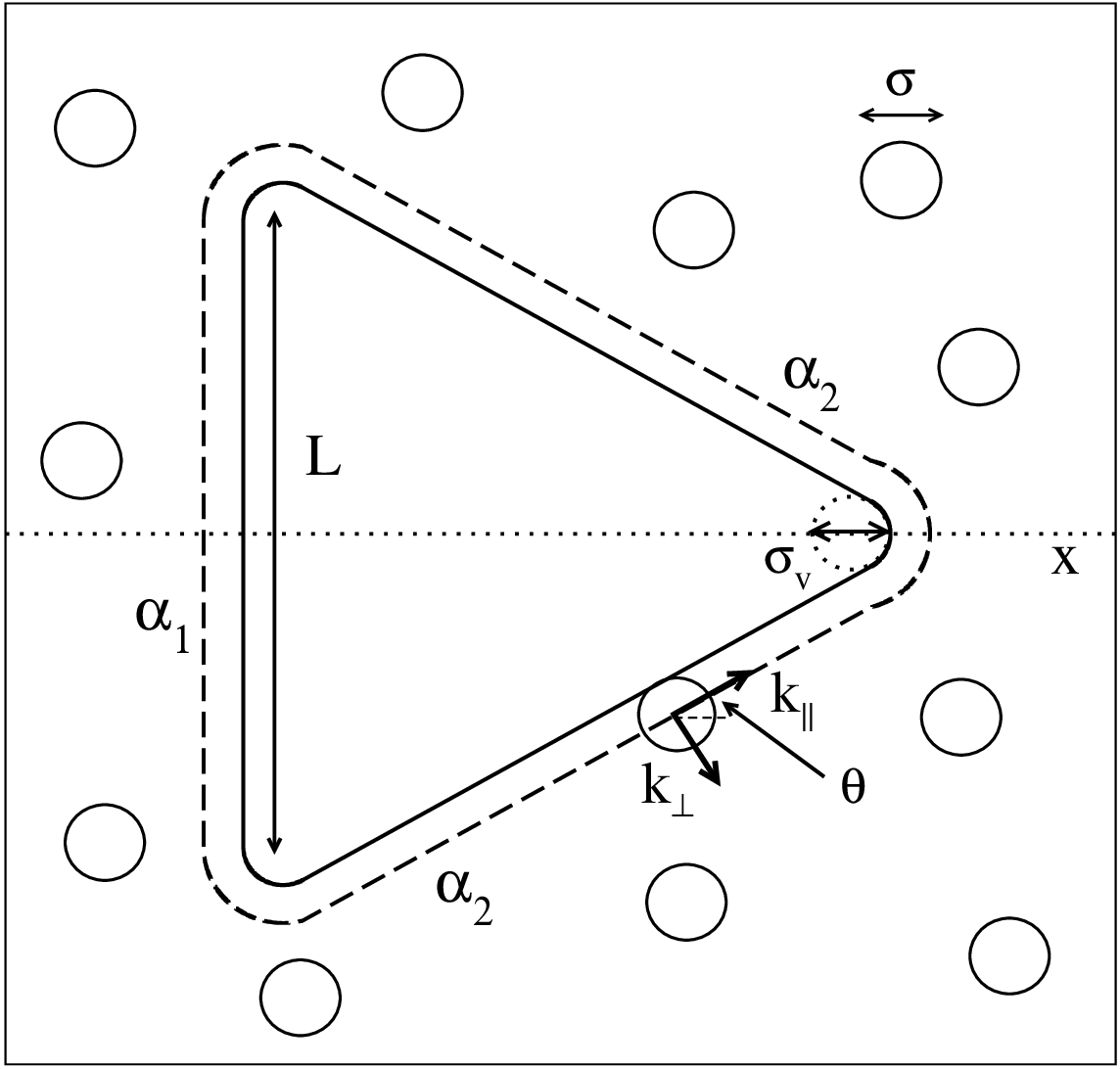}
\caption{Sketch of an inelastic equilateral triangle with different coefficient of restitution on left and right side ($\al_1$ and $\al_2$ respectively) and surrounding by a gas of inelastic particles. The dashed line correspond to the effective triangle seen by a colliding disk.
\label{triangle}}
\end{center}
\end{figure}

\begin{figure}[htbp]
\begin{center}
\includegraphics[angle=0,width=8.5cm,clip=true]{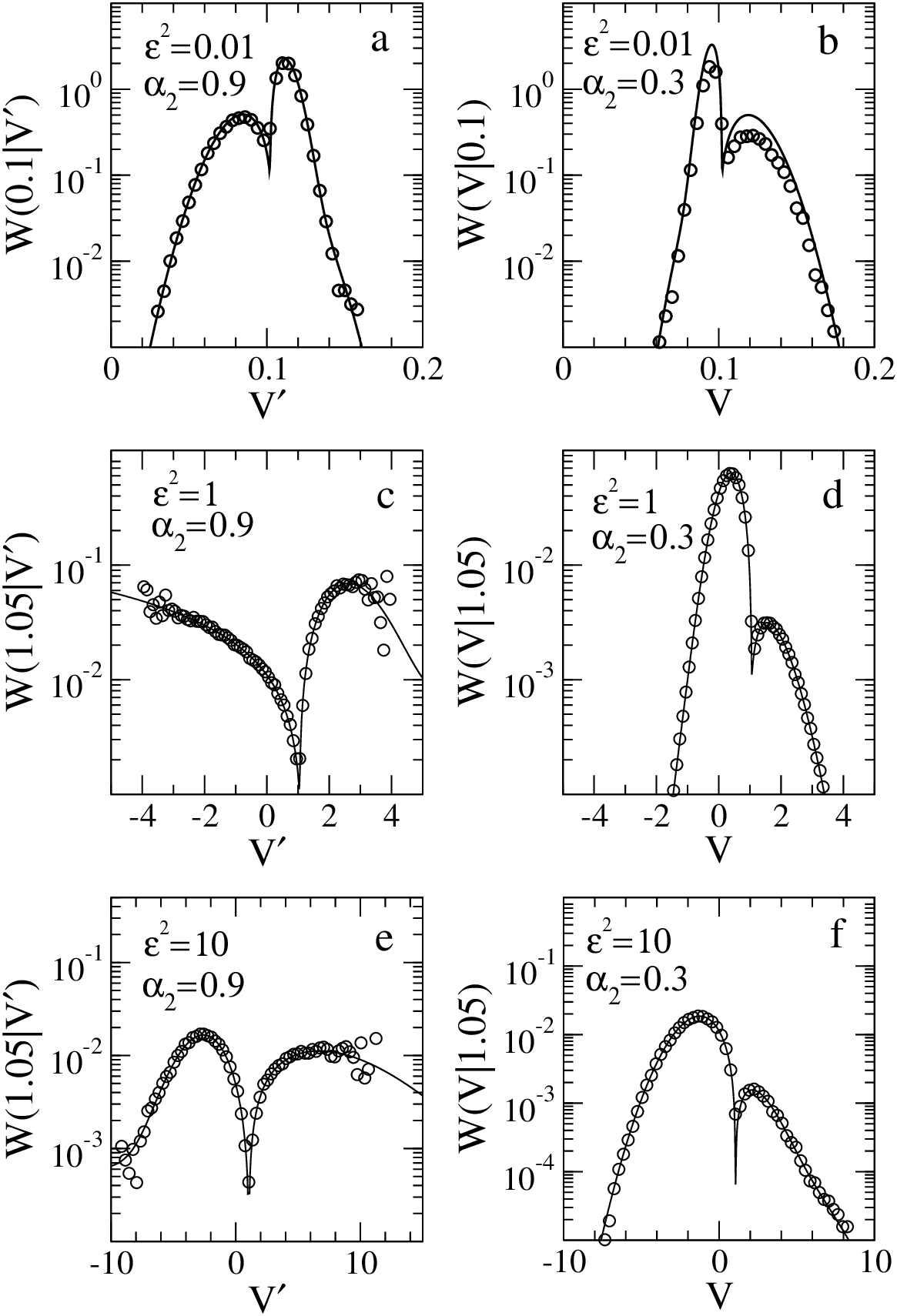}
\caption{The transition rate $W(V|V')$ of an inelastic equilateral triangle for different values of the mass ratio $\ep^2$ and the coefficient of restitution  $\al_2$. The panels a, c and e show the trend as a function of $V'$ for $\al_2=0.9$, $\al_{dd}=0.9$ and $\ep^2=0.01,1$ and $10$ respectively. The panels b, d and f instead show the trend as a function of $V$ for $\al_2=0.3$ and for the same values of $\ep^2$. The symbols correspond to the simulation data, while the lines are obtained from Eqs.(\ref{tr-rate}) and (\ref{gm1})-(\ref{gm2}).
\label{rates}}
\end{center}
\end{figure}

\begin{figure}[htbp]
\begin{center}
\includegraphics[angle=0,width=8.cm,clip=true]{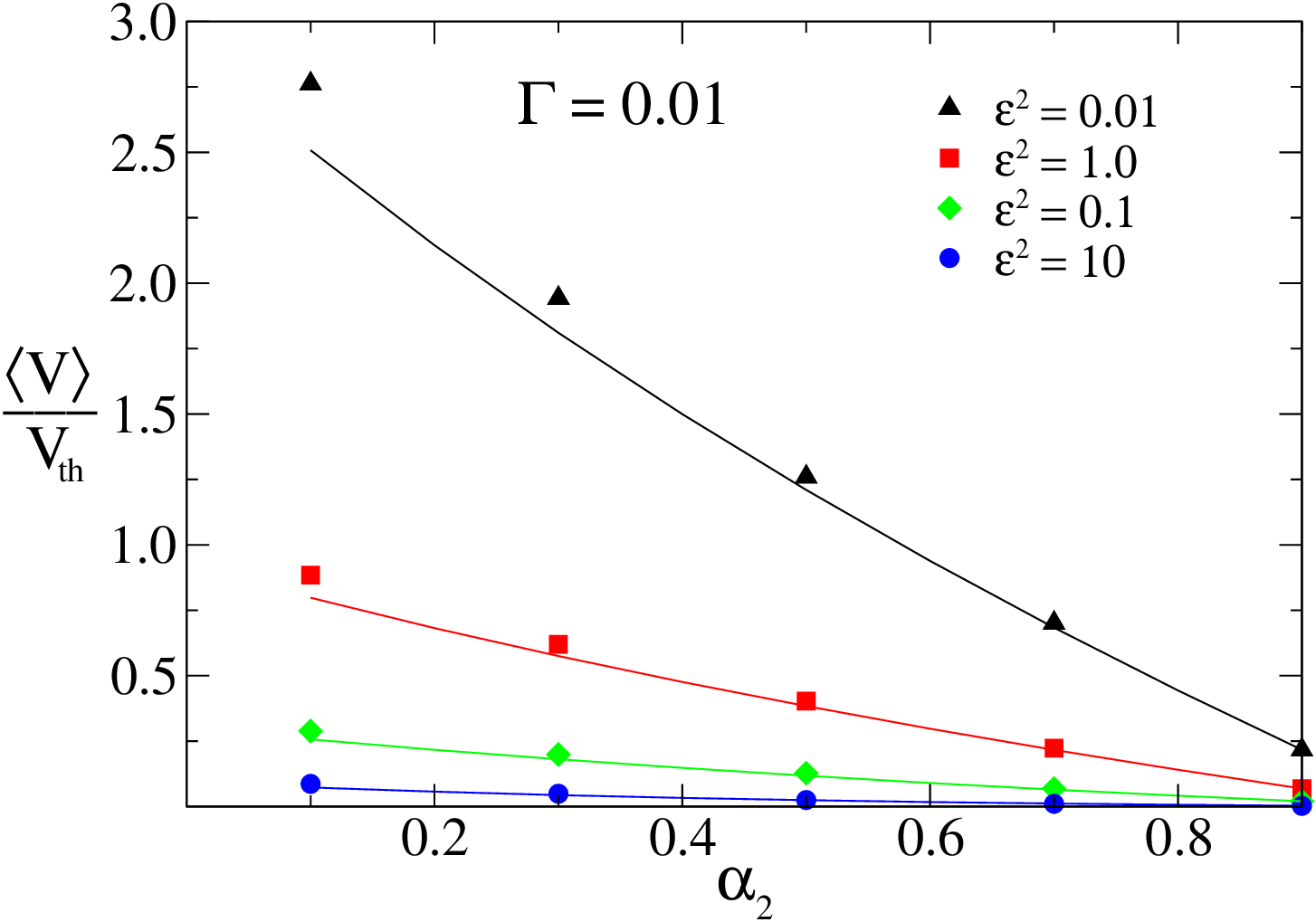}
\caption{The velocity of an equilateral triangle ratchet, rescaled with its thermal velocity $V_{th}=\sqrt{T_r/M}$, as function of the coefficient of restitution $\al_2$ for $\al_{dd}=0.9$, $\tau_b/\tau_{dd}=137.5$ and for different values of the parameter $\ep^2$: $10$ (circles), $1.0$ (squares), $0.1$ (diamonds) and $0.01$ (triangles). The symbols correspond to the simulation data while the lines correspond to the solutions obtained from Eqs.(\ref{avequations1})-(\ref{avequations6}).
\label{Vrat}}
\end{center}
\end{figure}

\begin{figure}[htbp]
\begin{center}
\includegraphics[angle=0,width=8.cm,clip=true]{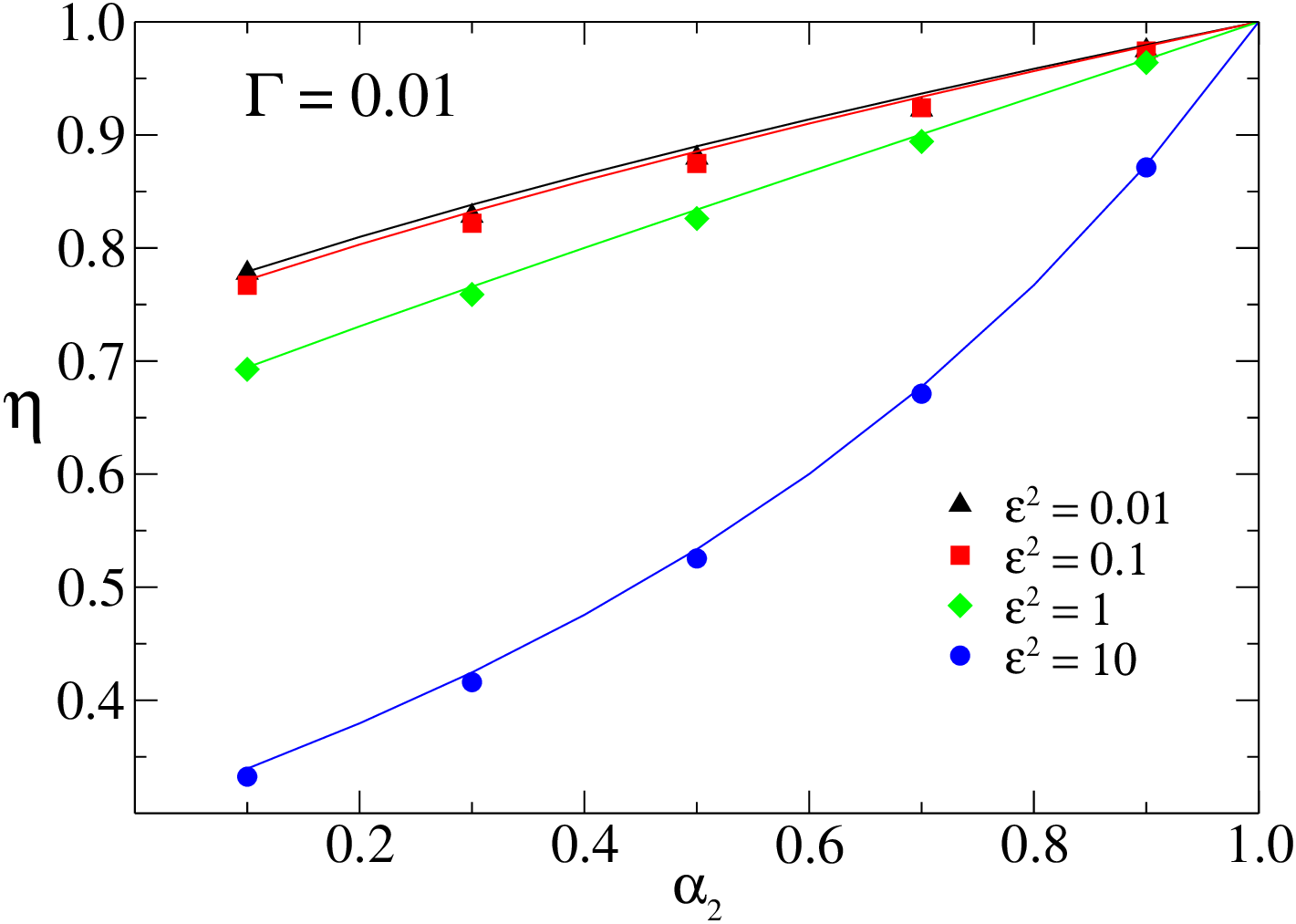}
\caption{The temperature ratio $\eta=T_r/T_g$ as function of the coefficient of restitution $\al_2$ for the same cases of Fig. \ref{Vrat}. The symbols correspond to the simulation data while the lines correspond to the solutions obtained from Eqs.(\ref{avequations1})-(\ref{avequations6}).
\label{eta}}
\end{center}
\end{figure}

The aim of this section is to compare our theoretical
results with DSMC numerical simulations, for a particular choice of
the ratcheting object. Note that the choice of DSMC algorithm always satisfies the Molecular Chaos. No constraints are imposed on the velocity 
pdfs: therefore this comparison is a test of
the many assumptions done about the pdfs, to obtain
Eqs. (\ref{avequations1})-(\ref{avequations6}).
For the sake of simplicity, we consider as object an equilateral triangle with
different inelasticities for the left face and for the two right faces
(see Fig.\ref{triangle}).  In particular, we choose that the coefficient of restitution 
is $\al_2$ if $ 0\leq
\theta \leq \pi$, and it is $\al_1$ if $\pi \leq \theta \leq 2\pi$.
The collision rules (\ref{eq:rules}) are well defined if the surface
is smooth. We consider then a triangle with the vertex shaped by
circular arches of radius $\sigma_v/2$. 
In this case the outer surface is $\tilde{C}=3L+\pi(\sigma+\sigma_v)$
where $L$ is the ``linear'' side of the triangle (see
Fig. \ref{triangle}). The function $F(\theta)$, using the symmetry of
the object, becomes
\begin{eqnarray}
F(\theta)=&\frac{L}{\tilde{C}}\Big[ \delta \Big(\theta-\frac{3}{2}\pi \Big) + 2\delta \Big(\theta-\frac{\pi}{6}\Big) \Big]+ \frac{r_e}{\tilde{C}}
\label{surf}
\end{eqnarray}
with $r_e=(\sigma+\sigma_v)/2$.\\
Moreover, the transition rate (\ref{trate}) can be written as
\begin{eqnarray}
W(V|V')=&n\tilde{C} \sqrt{\frac{m}{2\pi T_g}} \Big[(V'-V) \Theta(V'-V) g_{-}(V,V') + \nonumber  \\
&(V-V') \Theta(V-V')g_{+}(V,V')
\label{tr-rate}
\end{eqnarray}
where
\begin{eqnarray}
\label{gm}
g_{-}(V,V')&= \int_{\sen>0}d\theta F(\theta)\frac{\kappa^2(\theta,\al_2)}{\sens} \cdot  \nonumber \\
&\exp \Big\{-\frac{m\sens}{2T_g}\left[ \frac{\kappa(\theta,\al_2)}{\sens}(V-V')+V' \right]^2 \Big\} \equiv \nonumber \\
&\equiv\int_{\sen>0}d\theta F(\theta) ~\lambda(\theta,\al_2,V,V') \\
\label{gp}
g_{+}(V,V')&= \int_{\sen<0}d\theta F(\theta)~\lambda(\theta,\al_1,V,V').
\end{eqnarray}
The last line of Eq. (\ref{gm}) defines the function
$\lambda(\theta,\al(\theta),V,V')$. Using the Eq. (\ref{surf}) and the
symmetry respect to $\theta$, the above expressions become
\begin{eqnarray}
\label{gm1}
g_{-}(V,V')&= \frac{2L}{\tilde{C}}\lambda\left(\frac{\pi}{6},\al_2,V,V'\right)+
\frac{r_{e}}{\tilde{C}} \int_0^{\pi}\!\!\!\!d\theta~\lambda(\theta,\al_2,V,V') \\
g_{+}(V,V')&= \frac{L}{\tilde{C}}\lambda\left(\frac{\pi}{2},\al_1,V,V'\right)+
\frac{r_{e}}{\tilde{C}} \int_0^{\pi}\!\!\!\!d\theta ~\lambda(\theta,\al_1,V,V')
\label{gm2}
\end{eqnarray}
Some sections of the transition rate surface, for different values of
$\ep^2$, are shown in Fig. \ref{rates}, together with the simulation
data obtained from a DSMC with $\tau_b/\tau_{dd}=137.5$ and
$\al_{dd}=0.9$.
The Figure~\ref{rates} displays a very good agreement between theory
and DSMC for
the transition rate for all values of $\al_2$ and $\ep^2$ studied.\\
In Fig.~\ref{Vrat} we show the rescaled observable $\langle
V\rangle/V_{th}$ as a function of the right coefficient of restitution
$\al_2$ for different values of $\ep^2$. The theory results in good
agreement with the simulation data, supporting our assumptions. The
trends are analogue to those obtained in~\cite{epl}, where only the
inelasticity asymmetry was considered: such a similarity indicates that
the effects, due to the different inelasticity of the ratchet, are
predominant with respect to the geometrical asymmetry, that however can be not
neglected~\cite{noi}.  The data referring to ratio $\eta$ of the
granular temperatures of the system go in the same direction (see
Fig.~\ref{eta}).

\section{Conclusion}
\label{Conclusion}
Within the present article we have investigated the statistical
properties of a specific non equilibrium system, a 2D object sliding
along an axis and colliding inelastically with a granular gas coupled
to a thermal bath. If the object has an asymmetric shape or possesses
a non uniform inelasticity profile, one observes a net drift.  It is
possible, employing the BE, to describe theoretically the system. We
first obtain the collision frequency between ratchet and disks. Such a
quantity determines the interplay between the gas and the ratchet
dynamics. Secondly, we have derived a system of coupled equations of
some relevant averages of the distribution functions of the ratchet
and of the gas, respectively, describing the time evolution of the
whole system. In the limit of light objects ($\ep^2 \ll 1 $), it is
fundamental to include the third moment of the ratchet in order to
attain a satisfactory description of the ratchet behavior. We have
performed DSMC simulations in the case of an equilateral triangle and
compared the theoretical predictions with the numerical results and
found a fair agreement: in particular, we have measured the transition rate of
the tracer, its mean velocity and the granular temperature for difference
$\ep^2$ values. Whereas the theory gives an explicit representation of
such transition rates based on the assumption of a Maxwellian
velocity distribution for the gas particles, our simulations give a
direct estimate of the same quantity. The good agreement with the
analytic results confirms our assumptions.

\ack

The work of GS and AP is supported by the ``Granular-Chaos'' project,
funded by the Italian MIUR under the FIRB-IDEAS grant number RBID08Z9JE.


\appendix
\section*{Appendix}
The coefficients of the Eq.(\ref{collf}) and Eqs. (\ref{avequations1})-(\ref{avequations3}) can be written as
\begin{eqnarray}
\label{coeff}
\fl u_1(z)=\int_0^{2\pi}d\theta F(\theta)\sqrt{z+\sens} \nonumber \\
\fl u_2(z)=\int_0^{2\pi}d\theta F(\theta)\frac{\senf}{(z+\sens)^{3/2}} \nonumber \\
\fl u_3(z)=\int_0^{2\pi}d\theta F(\theta)\frac{\sin^3\theta \cs}{(z+\sens)^{3/2}} \nonumber \\
\fl a_1(z)=\int_0^{2\pi}d\theta F(\theta) \frac{\sen}{1+\ep^2\sens}  (1+\al(\theta))\left(z+\sens\right)\nonumber \\
\fl a_2(z)=\int_0^{2\pi}d\theta F(\theta) \frac{\sens}{1+\ep^2\sens} (1+\al(\theta))\sqrt{z+\sens}\nonumber \\
\fl a_3(z)=\int_0^{2\pi}d\theta F(\theta) \frac{\sen\cs}{1+\ep^2\sens} (1+\al(\theta))\sqrt{z+\sens}\nonumber \\
\fl a_4(z)=\int_0^{2\pi}d\theta F(\theta) \frac{(1+\al(\theta))\senf}{ (1+\ep^2\sens) \sqrt{z+\sens}}\nonumber \\
\fl b_1(z)=\int_0^{2\pi}d\theta F(\theta) \frac{(1+\al(\theta))\ep^4\sens} {(1+\ep^2\sens)^2}\sqrt{z+\sens}  \left[z(1+\al(\theta))+\sens(\al(\theta)-1)-2\right] \nonumber \\
\fl b_2(z)=\int_0^{2\pi}d\theta F(\theta) \frac{\sen} {(1+\ep^2\sens)^2}
\Big\{ 3(1+\al(\theta))^2 z \ep^4\sens+ \left[1-(2+3\al(\theta))\ep^2\sens\right]\cdot \nonumber \\ (1-\al(\theta)\ep^2\sens)\Big\} \nonumber \\
\fl b_3(z)=\int_0^{2\pi}d\theta F(\theta) \frac{\cs} {(1+\ep^2\sens)^2}
\Big\{ 3(1+\al(\theta))^2 z \ep^4\sens+ \left[1-(2+3\al(\theta))\ep^2\sens\right] \cdot \nonumber \\ (1-\al(\theta)\ep^2\sens)\Big\} \nonumber \\
\fl b_4=\int_0^{2\pi}d\theta F(\theta) \frac{\sen}{(1+\ep^2\sens)^2} \left(1-\al(\theta)\ep^2\sens\right)^2 \nonumber \\
\fl b_5(z)=\int_0^{2\pi}d\theta F(\theta) \frac{\sens}{(1+\ep^2\sens)^2 (z+\sens)^{5/2}} \Big\{ 6z^2 (1-\al(\theta)\ep^2\sens)^2+z\sens \cdot \nonumber \\
\Big[ 12(1+\al(\theta))^2\ep^4\senf + 
5(1+\ep^2\sens)\left(1+\ep^2\sens-4(1+\al(\theta))\ep^2\sens\right) \Big] \nonumber \\ +2\senf (1-\al(\theta)\ep^2\sens)\Big[1-(2+3\al(\theta))\ep^2\sens\Big]- 
(1+\ep^2\sens)^2 \cdot \nonumber \\
(6 z^2+5z\sens+2\senf)\Big\}\nonumber \\
\fl b_6(z)=\int_0^{2\pi}d\theta F(\theta) \frac{\sen\cs}{(1+\ep^2\sens)^2 (z+\sens)^{5/2}} \Big\{ 6z^2 (1-\al(\theta)\ep^2\sens)^2+z\sens \cdot \nonumber \\
\Big[ 12(1+\al(\theta))^2\ep^4\senf + 
5(1+\ep^2\sens)\left(1+\ep^2\sens-4(1+\al(\theta))\ep^2\sens\right) \Big] \nonumber \\ +2\senf (1-\al(\theta)\ep^2\sens)\Big[1-(2+3\al(\theta))\ep^2\sens\Big]-
(1+\ep^2\sens)^2 \cdot \nonumber \\
(6 z^2+5z\sens+2\senf)\Big\}\nonumber \\
\fl c_1(z)=\int_0^{2\pi}d\theta F(\theta)  \frac{\sen}{(1+\ep^2\sens)^3} \Big\{ (1+\al(\theta))^3 z^2 \ep^6\sens + (1+\al(\theta))z\ep^2 \Big[1-(1+2\al(\theta)) \cdot \nonumber \\
\ep^2\sens \Big] 
(1-\al(\theta)\ep^2\sens) -(1-\al(\theta)\ep^2\sens)^3\Big\} \nonumber \\
\fl c_2(z)=\int_0^{2\pi}d\theta F(\theta) \frac{\sens}{(1+\ep^2\sens)^3 (z+\sens)^{3/2}} \Big\{
8z^3\ep^6\sens(1+\al(\theta))^3+6z^2\ep^2 \Big [1+\al(\theta) - \nonumber \\
(1+\al(\theta))(1+3\al(\theta))\ep^2\sens + 
(1+\al(\theta))(2+5\al(\theta)+4\al^2(\theta))\ep^4\senf \Big] + \nonumber \\
 z \Big[ (1+2\al(\theta))\ep^2\sens-1 \Big]^3 + 2\sens 
(1-\al(\theta)\ep^2\sens)  [(3+4\al(\theta))\ep^2\sens -\nonumber \\
1] - (1+\ep^2\sens)^3\left(3z+2\sens \right) \Big\} \nonumber \\
\fl c_3(z)=\int_0^{2\pi}d\theta F(\theta)  \frac{1}{(1+\ep^2\sens)^3 (z+\sens)^{5/2}} \Big \{ 2z^3(1-\al(\theta)\ep^2\sens)^2 [1-(3+4\al(\theta))\cdot \nonumber \\
\ep^2\sens] + z^2\sens \Big[ 15-5(1+10\al(\theta))\ep^2\sens+5(1+4\al(\theta) - 12\al^2(\theta)) \cdot \nonumber \\
\ep^4\senf+(1-2\al(\theta)-12\al^2(\theta)-24\al^3(\theta))\ep^6\sensix \Big] + 2z\senf \Big[ 10 -(1+ \nonumber \\
31\al(\theta)) 
\ep^2\sens +(1+4\al(\theta)- 33\al^2(\theta))\ep^4\senf- \al(\theta)(1+3\al(\theta)+12\al^2(\theta))\cdot\nonumber \\
\ep^6\sensix \Big] + 8\sensix 
(1-\al(\theta)\ep^2\sens)^3 - (1+\ep^2\sens)^3 \Big[2z^3+15z^2\sens+ \nonumber \\
 20z\senf+8\sensix \Big] \Big\} \nonumber \\
\fl c_4=\int_0^{2\pi}d\theta F(\theta)  \frac{\sen}{(1+\ep^2\sens)^3}(1-\al(\theta)\ep^2\sens)^2 \left[ 1-(3+4\al(\theta))\ep^2\sens  \right] \nonumber \\
\fl c_5(z)=\int_0^{2\pi}d\theta F(\theta) \frac{\ep^2\sens}{(1+\ep^2\sens)^3 \left(z+\sens \right)^{3/2}} \Big \{ 8z^3(1+\al(\theta))^3\sens + 6z^2(1+\al(\theta)) \cdot \nonumber \\
\Big[ 1-(1+3\al(\theta))\ep^2\sens +
(2+5\al(\theta)+4\al^2(\theta))\ep^4\senf \Big] +6z(1+\al(\theta))\sens \cdot \nonumber \\
\Big[ 3-6\al(\theta)\ep^2\sens + 
(1+2\al(\theta)+4\al^2(\theta)\ep^4\senf \Big] +2\senf 
\Big[ 6(1+\al(\theta)) +\nonumber \\ 
3(1+\al(\theta))(1-3\al(\theta))\ep^2\sens + (1+3\al^2(\theta)+4\al^3(\theta)) \Big] \Big\} \nonumber \\
\fl c_6(z)=\int_0^{2\pi}d\theta F(\theta) \frac{\ep^2\sen\cs}{(1+\ep^2\sens)^3 \left(z+\sens \right)^{3/2}} \Big \{ 8z^3(1+\al(\theta))^3\sens + 6z^2(1+\al(\theta))\nonumber \\
 \Big[ 1-(1+3\al(\theta))\ep^2\sens + 
(2+5\al(\theta)+4\al^2(\theta))\ep^4\senf \Big] +6z(1+\al(\theta))\sens \cdot \nonumber \\ 
\Big[ 3-6\al(\theta)\ep^2\sens + 
(1+2\al(\theta)+4\al^2(\theta))\ep^4\senf \Big] +2\senf 
\Big[ 6(1+\al(\theta)) + \nonumber \\
3(1+\al(\theta))(1-3\al(\theta))\ep^2\sens + (1+3\al^2(\theta)+4\al^3(\theta)) \Big] \Big\} \nonumber \\
\fl c_7=\int_0^{2\pi}d\theta F(\theta) \frac{\sen}{(1+\ep^2\sens)^3} \Big[ 1-3(1+2\al(\theta))\ep^2\sens+3\al(\theta)(2+3\al(\theta))\ep^4\senf- \nonumber \\
\al^2(\theta) (3+4\al(\theta))\ep^6\sensix \Big] \nonumber \\
\fl c_8=\int_0^{2\pi}d\theta F(\theta) \frac{\cs}{(1+\ep^2\sens)^3} \Big[ 1-3(1+2\al(\theta))\ep^2\sens+3\al(\theta)(2+3\al(\theta))\ep^4\senf- \nonumber \\
\al^2(\theta)(3+4\al(\theta))\ep^6\sensix \Big] \nonumber \\
\fl d_1(z)=\int_0^{2\pi}d\theta F(\theta) \frac{\cs}{1+\ep^2\sens}  (1+\al(\theta))\left(z+\sens\right) \nonumber \\
\fl d_2(z)=\int_0^{2\pi}d\theta F(\theta) \frac{(1+\al(\theta))\css}{1+\ep^2\sens} \sqrt{z+\sens} \nonumber \\
\fl d_3(z)=\int_0^{2\pi}d\theta F(\theta) \frac{(1+\al(\theta)) \sin^3\theta\cs}{ (1+\ep^2\sens) \sqrt{z+\sens}} \nonumber \\
\fl e_1(z)=\int_0^{2\pi}d\theta F(\theta) \frac{1+\al(\theta)}{1+\ep^2\sens} \sqrt{z+\sens} \left[z(\al(\theta)-1-2\ep^2\sens)+(1+\al(\theta))\sens \right]  \nonumber \\
\fl e_2(z)=\int_0^{2\pi}d\theta F(\theta) \frac{\sen}{(1+\ep^2\sens)^2} \Big \{ 3(1+\al(\theta))^2\sens+z \Big[ 2(1+\ep^2\sens)^2+(1+\al(\theta))\cdot \nonumber \\
(3\al(\theta)-1-4\ep^2\sens) \Big] \Big\} \nonumber \\
\fl e_3(z)=\int_0^{2\pi}d\theta F(\theta) \frac{\cs}{(1+\ep^2\sens)^2} \Big \{ 3(1+\al(\theta))^2\sens+z \Big[ 2(1+\ep^2\sens)^2+(1+\al(\theta))\cdot \nonumber \\
(3\al(\theta)-1-4\ep^2\sens) \Big] \Big\} \nonumber \\
\fl e_4=\int_0^{2\pi}d\theta F(\theta) \frac{(1+\al(\theta))^2\sin^3\theta}{(1+\ep^2\sens)^2} \nonumber \\
\fl e_5(z)=\int_0^{2\pi}d\theta F(\theta) \frac{(1+\al(\theta))\senf} {(1+\ep^2\sens)^2(z+\sens)^{5/2}} \Big[ z^2(5+3\al(\theta)+2\ep^2\sens) + 2z\sens\cdot \nonumber \\
(4+3\al(\theta)+\ep^2\sens) + 
3(1+\al(\theta))\senf \Big]\nonumber \\
\fl e_6(z)=\int_0^{2\pi}d\theta F(\theta) \frac{(1+\al(\theta))\sin^3\theta\cs} {(1+\ep^2\sens)^2(z+\sens)^{5/2}} \Big[ z^2(5+3\al(\theta)+2\ep^2\sens) + 2z\sens
\cdot \nonumber \\
(4+3\al(\theta)+\ep^2\sens) + 
3(1+\al(\theta))\senf \Big] \nonumber
\end{eqnarray}


\end{document}